\documentclass[12pt,preprint]{aastex}

\begin{document}

\title{The Progenitor of the Type Ia Supernova that created SNR 0519-69.0 in the Large Magellanic Cloud}
\author{Zachary I. Edwards}
\affil{Department of Earth and Space Sciences, Columbus State University, Columbus, GA 31907}
\email{Edwards\_Zachary@columbusstate.edu}

\and

\author{Ashley Pagnotta, Bradley E. Schaefer}
\affil{Department of Physics and Astronomy, Louisiana State University, Baton Rouge, LA 70803}

\begin{abstract}

Models for the progenitor systems of Type Ia supernovae can be divided into double-degenerate systems, which contain two white dwarfs, and single-degenerate systems, which contain one white dwarf plus one companion star (either a red giant, a subgiant, or a $>$1.16 M$_{\odot}$ main sequence star). The white dwarf is destroyed in the supernova explosion, but any non-degenerate companion remains intact.  We present the results of a search for an ex-companion star in SNR 0519-69.0, located in the Large Magellanic Cloud, based on images taken with the {\it Hubble Space Telescope} with a limiting magnitude of $V=26.05$. SNR 0519-69.0 is confidently known to be from a Type Ia supernova based on its light echoes and X-ray spectra. The geometric center of the remnant (based on the H$\alpha$ and X-ray shell) is at 05:19:34.83, -69:02:06.92 (J2000).  Accounting for the measurement uncertainties, the orbital velocity, and the kick velocity, any ex-companion star must be within 4.7$\arcsec$ of this position at the 99.73\% confidence level.  This circle contains 27 main sequence stars brighter than $V=22.7$, any one of which could be the ex-companion star left over from a supersoft source progenitor system.  The circle contains no post-main sequence stars, and this rules out the possibility of all other published single-degenerate progenitor classes (including symbiotic stars, recurrent novae, helium donors, and the spin-up/spin-down models) for this particular supernova.  The only remaining possibility is that SNR 0519-69.0 was formed from either a supersoft source or a double-degenerate progenitor system.

\end{abstract}


\section{Introduction}

Type Ia supernovae (SNe Ia) are a critical component of many areas of modern astrophysics, including stellar evolution and cosmology. It is widely accepted that the detonation is caused by the thermonuclear explosion of a carbon-oxygen white dwarf (WD) that has reached the Chandresekhar mass limit.  A variety of stellar systems have been proposed as progenitors.  These progenitor candidates can be divided into two classes, the double-degenerates, DD (Tutukov \& Yungelson 1981; Van Kerkwijk et al. 2010), and the single-degenerates, SD (Iben \& Tutukov 1984; Whelan \& Iben 1973), based whether the system has one or two WDs.  After considering the full array of WD binaries (Branch et al. 1995; Parthasarathy et al. 2007), reasonable SD progenitor models include the recurrent novae, symbiotic stars, supersoft sources, helium stars, and spin-up/spin-down systems (Hachisu \& Kato 2001; Hachisu et al. 1999a; Hachisu et al. 1999b; Langer et al. 2000; Wang et al. 2009; Justham 2011).  There is plausible evidence that the observed SNe Ia events could arise from multiple progenitor channels (Maoz \& Mannucci 2011). 

For decades the progenitor question has gone unanswered; over the last decade the use of SNe Ia as cosmology tools has elevated the importance of this problem (c.f. Blandford et al. 2010).  The progenitor problem can be approached from many directions.  The recent nearby supernova SN 2011fe has had modest limits placed (Li et al. 2011) on its pre-eruption progenitor magnitude ($M_V$ must be fainter than -1), but this limit can only reject the most luminous red giant companion stars in a symbiotic system.  Limits based on death rates of progenitor candidates from population synthesis models have uncertainties that are too large to have any real utility in deciding between candidates.  The recent claim to eliminate all SD models (Gilfanov \& Bogdan 2010), based on the supersoft X-ray flux from elliptical galaxies, has been broadly rejected for many strong reasons (Meng \& Yang 2011; Di Stefano 2010; Hachisu et al. 2010; Lipunov et al. 2010; Orio et al. 2010).  The lack of even a small amount of hydrogen in SN Ia spectra (e.g. Leonard 2007) would appear to reject all SD candidates, but this result is ambiguous because detailed models show that the hydrogen will not be visible (Marietta et al. 2000), and some events have been seen to have hydrogen (Branch et al. 1983; Hamuy et al. 2003; but see Livio \& Riess 2003).  In all, there has been no decisive evidence proving or disproving any one progenitor candidate or class, and our community is roughly evenly divided in opinions.

A promising method (Ruiz-Lapuente 1997; Canal et al. 2001) to distinguish between candidates is to look for the former companion star near the center of nearby supernova remnants (SNRs), because the various progenitor classes have different types of companion stars that survive the explosion.  Symbiotic progenitors must leave behind a red giant star in the middle of the SNR, while a helium star progenitor must leave the luminous helium star near the SNR center.  A recurrent nova progenitor must leave either a red giant or subgiant.  (Some recurrent novae have main sequence companion stars, but these systems are dominated by the prior classical nova event that keeps the WD from gaining mass over each cycle; Schaefer et al. 2010.)  Supersoft sources must have ex-companions that are either subgiants or main sequence stars with mass $>$1.16 $M_{\odot}$ (Schaefer \& Pagnotta 2012). This is the minimum mass for which the system, with its near-Chandrasekhar mass WD, will have a mass ratio of $\geq 5/6$, which is required to drive fast accretion via Roche lobe overflow onto the surface of the WD, and is confirmed by detailed models (Langer et al. 2000). Published spin-up/spin-down models (e.g. Justham 2011) account for angular momentum conservation in SD systems, spinning up the WD as it accretes, which delays the explosion trigger. After the cessation of accretion, the WD spins down until the supernova is triggered. Due to short time limits on the spin-down time (Yoon \& Langer 2005), the companion star will still be either a bright red giant or subgiant core. Importantly, these ex-companion stars will not change their luminosity or surface temperature greatly; they should appear at largely the same place on the HR diagram several centuries after the explosion (Marietta et al. 2000; Pan et al. 2010; Podsiadlowski 2003).  The DD model predicts that there will be no ex-companion star.  Thus, by looking near the center of a Type Ia SNR, the existence and nature of any ex-companion star will distinguish the progenitor system.

This method can only be applied to SNRs which definitely came from SNe Ia.  In our galaxy, only Tycho's SNR (SN 1572) and the remnant from SN 1006 are confidently known to be from SNe Ia (Schaefer 1996; Krause et al. 2008).  Ruiz-Lapuente et al. (2004) looked near the center of the Tycho remnant and identified a G2 IV star (i.e., a subgiant) called ÔStar GÕ near the center as the ex-companion based on its high proper motion at the right distance.  If this identification is confirmed, it would immediately rule out the DD and symbiotic channels, pointing to the recurrent nova path.  Unfortunately, various properties are still being disputed (Ihara et al. 2007; Kerzendorf et al. 2009; Hernandez et al. 2009).  Hernandez et al. (2009) performed a detailed analysis of hundreds of metal absorption lines to find that nickel and cobalt are anomalously over-abundant in the atmosphere of Star G.  If confirmed, this would strongly point to Star G having contamination from SN ejecta.  For now, the case is unresolved, although we are inclined to think that Star G is the ex-companion star based on the Hernandez et al. (2009) paper.  For the case of SN 1006, a red giant ex-companion star can be ruled out (Ruiz-Lapuente 2011).  A major challenge is that galactic SNRs have large uncertainties in distances, high extinction, and crowded fields.

Recently, we have applied this same method to SNRs in the Large Magellanic Cloud (LMC).  Four LMC SNRs have been confidently identified as coming from SNe Ia based on light echo spectroscopy (Rest et al. 2005; 2008; A. Rest 2010, private communication) and X-ray spectroscopy (Hughes et al. 1995).  The youngest and most symmetric of these is SNR 0509-67.5, which is from a SN 1991T class Type Ia explosion that occurred 400$\pm$50 years ago.  The {\it Hubble Space Telescope} (HST) has taken deep images of SNR 0509-67.5 in B, V, I, and H$\alpha$.  Schaefer \& Pagnotta (2012) report that the maximal (99.73\% containment) central error circle is empty of point sources to V=26.9, which corresponds to $M_V$=+8.4 mag in the LMC.  They also show that all SD models require the ex-companion to be more luminous than $M_V$=+4.2 mag (V=22.7 in the LMC).  With this, all published SD models are rejected with high confidence.  The only remaining model is the DD, so the strong conclusion is that the SN Ia event leading to SNR 0509-67.5 came from a double-degenerate system.

In this paper, we consider a second LMC remnant, SNR 0519-69.0.  The light echo shows the supernova spectrum to be that of a normal SN Ia, with an age of 600$\pm$200 years (Rest et al. 2005; A. Rest 2010, private communication).  We have used the same techniques as in Schaefer \& Pagnotta (2012) on SNR 0519-69.0, applied to archival V and H$\alpha$ HST images.  The case of SNR 0519-69.0 is not as optimal as for SNR 0509-67.5, because the SNR is older and more irregular in outline (leading to a substantially larger central error ellipse) and the star density is higher (so the error circle is not empty of stars).  The next three sections detail the size and position of the central error circle, the contents of that region, and the severe constraints on any possible progenitor model.  In short, we demonstrate that SNR 0519-69.0 does not have any post-main sequence ex-companion star, and this rejects the symbiotic, recurrent nova, helium star, and spin-up/spin-down SD models.

\section{The Center of SNR 0519-69.0}

We have used public domain images of SNR 0519-69.0 from both HST and from the {\it Chandra} X-ray observatory.  The HST images were taken with the Advanced Camera for Surveys camera in April 2010 (J. P. Hughes P.I.). F658N (H$\alpha$) and F550M (V band) observations were taken for a total of 4757 and 750 seconds, respectively.  The data were processed and combined using the standard PyRAF procedures and were analyzed using the IRAF {\it phot} package.  The combined H$\alpha$ and V image is presented in Figure 1.  The {\it Chandra} X-ray observations in June 2000 had a total exposure of 40.6 kiloseconds (S. S. Holt P.I.), with images available in three energy bands: 0.3-0.72 keV, 0.72-1.05 keV, 1.05-10 keV (Williams et al. (2001)).

The geometric center of the SNR was measured by two methods, both using the same procedure, but operating off different data sets corresponding to gas at greatly different temperatures.  We constructed nine sets of perpendicular bisectors from edge to edge across the remnant, each tilted approximately 10$\degr$ from the previous set. The center measurement from each set was retained and averaged together to obtain the geometric center. The RMS scatter of the nine measurements is a good estimate of the measurement uncertainty. Table 1 presents the geometric centers for both the H$\alpha$ shell and the X-ray shell.  These two positions differ by 0.5$\arcsec$ from the average, which is a measure of the systematic uncertainty from any one position by this method.  The two centers have similar measurement uncertainty, and we know of no reason to prefer one over the other, so the overall best estimate of the geometric center is taken to be a straight average of the two positions (see Table 1).  The uncertainty in this best geometric center comes from the addition in quadrature of the measurement error (0.7$\arcsec$) and the systematic uncertainty (0.5$\arcsec$), for a total of 0.9$\arcsec$.

The offset from the geometric center to the explosion site arises from asymmetries in the ejection velocity and the surrounding interstellar medium.  Schaefer \& Pagnotta (2012) note that the ejection velocity asymmetries are expected to be small, less than 10\%, while any bipolar component of the ejection will produce zero offset.  Asymmetric bubbles or clumps of gas or dust surrounding the expanding shell will produce an offset.  In all of these cases, the mechanism that causes the offset will also deform the edges of the shell, causing an out-of-round axis ratio.  Schaefer \& Pagnotta (2012) derive the offset from the measured axis ratio, but unfortunately the {\it direction} of the offset depends on knowing the cause of the deformation.  For SNR0519-69.0, the shell is closely round with no apparent deformation.  The edges of the shell display minor small-scale bumps, but such will lead to no significant offset.  The {\it Spitzer} 24-micron infrared image shows the swept up and heated dust grains (Williams et al. 2011), with three broad brightenings evenly distributed around the edge of SNR 0519-69.0, showing no apparent association with asymmetries, so it appears that there are no systematic variations that deform one quadrant of the shell or cause any significant offset.  A substantial qualification on this statement is that one quadrant of the remnant (towards the northeast), has a faint and thin outer arc, with a much brighter inner arc that looks like it has run into some relatively dense region of the gas and dust.  The outer edge, however, is closely circular, so the best evidence is that the outer edge is not affected by either type of asymmetry, hence there is a near zero offset between the geometric center and the site of the explosion. This conclusion will have some 1$\sigma$ uncertainty attached to it, and we take this to be close to the RMS variation in the outer radius of the remnant, which is 5\% of the radius of the SNR (i.e., 0.9$\arcsec$).  When combined with the measured position of the geometric center, we get the position of the explosion with an uncertainty of 0.9$\arcsec$ added in quadrature with 0.9$\arcsec$, to get a radius for the error circle of 1.3$\arcsec$. 

The position of any ex-companion star will be offset from the explosion site due to the proper motion of the star.  The orbital velocity of the companion will depend on its mass and size (because it is filling its Roche lobe).  The kick velocity onto the companion will be perpendicular to the orbital velocity, and will depend substantially on the size and closeness of the companion.  These effects have been calculated by Marietta et al. (2000), Canal et al. (2001), Pan et al. (2010), and Schaefer \& Pagnotta (2012).  For red giant and subgiant companions, the orbital velocities and kick velocities will be relatively small, 100 and 250 km s$^{-1}$ respectively.  For main sequence companion stars, it is critical to realize that all models require the star to be more massive than 1.16 M$_{\odot}$, as this is the limit required to drive fast accretion onto the white dwarf (Schaefer \& Pagnotta 2012; Langer et al. 2000).  With this limit on the mass, the orbital velocity (relative to the WD) plus the kick velocity require the relative space velocity of any main sequence progenitor companion to be 390 km s$^{-1}$ or less.  For a distance modulus to the LMC of 18.50$\pm$0.10 mag (Freedman et al. 2001; Schaefer 2008), a velocity of 390 km s$^{-1}$ over a time interval of 600$\pm$200 years corresponds to a maximum proper motion of 1.0$\pm$0.3$\arcsec$.  The proper motion will be 0.25$\pm$0.08$\arcsec$ and 0.6$\pm$0.2$\arcsec$ for red giants and subgiants, respectively.

The size of the final error circle depends on the class of the companion and the confidence level required to be contained within the circle.  The distribution of the uncertainties for the geometric center and the offset to the explosion site are Gaussian, but the distribution for the proper motion is edge dominated, so a convolution is needed to express the final distribution for the position of any ex-companion.  We report the error circle radii for which 99.73\% (i.e., 3$\sigma$) of the ex-companion stars will be contained.  We calculate that for the minimal mass main sequence star, the 99.73\% error radius is 4.7$\arcsec$.  For subgiants and red giants, the error circles are 4.5$\arcsec$ and 4.3$\arcsec$ respectively.

\section{Contents of the Central Region}

We have just two bands for our HST images, V (F550M) and H$\alpha$ (F658N).  For each of the stars in the two images, we have performed aperture photometry with the standard HST zeros.  The limiting magnitude is V=26.05 to the 5$\sigma$ detection threshold.  We calculate a non-standard color, $V-H\alpha$.  For all 127 stars within the 4.7$\arcsec$ error radius, we have tabulated the HST astrometry and photometry in Table 2.  We have added two stars of interest outside the error circle.  Table 2 is presented in the printed journal with only the top ten lines displayed, so we have placed the brightest star in the circle (ID=1), the nearest red giant (ID=RG, outside the circle), and the nearest possible subgiant star (ID=SG, outside the circle) at the top of the list, with the remainder of the stars listed in order of their angular distance from the center ($\Theta$). The full table is available in the online version of this manuscript.

We have constructed a color-magnitude diagram with the V and $V-H\alpha$ magnitudes.  This can be converted to an HR diagram with the distance modulus to the LMC of 18.50$\pm$0.10.  Unfortunately, this HR diagram uses nonstandard colors, as this is all that is currently available from HST.  The wavelength separation between the two bands is not as large as desired, and this makes for relatively small separation between subgiants and main sequence stars.  The exposure times (12.5 minutes in V and 79 minutes in H$\alpha$) along with the narrow bandpass for H$\alpha$ make for relatively poor photon statistics at the faint end.  Nevertheless, the HST HR diagram for SNR 0519-69.0 shows a clear main sequence, red clump, and red giant branch.  This allows for simple identification of each star by its luminosity class.  As always, there are difficulties with distinguishing subgiant stars positioned near the main sequence, and we expect that ordinary measurement errors will shift some of the many main sequence stars out towards the subgiant region on the HR diagram.

The contents of the central region include 127 stars, of which 27 are main sequence stars brighter than V=22.7.  Schaefer \& Pagnotta (2012) have pointed out that such stars could be ex-companion stars for the supersoft source progenitor model.  Thus, there are 27 candidate ex-companion stars. The other 100 stars in the central region are main sequence stars that are fainter than the required V=22.7, therefore not possible ex-companions. Importantly, the stars in the central region do not include any post-main-sequence stars. That is, there are no red giants or subgiants within the 4.7$\arcsec$ error circle.  The nearest red giant and subgiant are 6.0$\arcsec$ and 7.4$\arcsec$ from the center, respectively.  There is no chance for any progenitor model that has a red giant or subgiant donor star.

\section{Conclusions}

The detection or exclusion of ex-companion stars near the center of Type Ia supernova remnants can provide decisive information about the progenitor.  For SNR 0519-69.0, the 99.73\% error circle contains no post-main sequence stars, which eliminates the symbiotic, recurrent nova, helium star, and spin-up/spin-down progenitor models.  Among the published single-degenerate models, only the supersoft source model is possible for SNR 0519-69.0.  A double-degenerate system has no ex-companion star, so this model is also fully consistent with our images.  Thus, for SNR 0519-69.0, the only possible progenitors are a supersoft source or a double-degenerate.

We note that Badenes et al. (2007) have examined SNR 0519-69.0 for evidence of a large cavity in the circumstellar region that would be carved out by fast, optically thick outflows that occur in the accretion wind scenario proposed for supersoft sources (Hachisu et al. 1996). They found that the dynamics of the ejecta from SNR 0519-69.0 are inconsistent with the predictions. If the accretion wind scenario is in fact the only possible method for obtaining a Ia from a supersoft source, then the lack of a cavity excludes the supersoft source progenitor class for SNR 0519-69.0 and we would be left with the double-degenerate model as the only viable progenitor. Like SNR 0509-67.5, which has been shown to be a double-degenerate, SNR 0519-69.0 also comes from a region with a low star formation history, which slightly favors a delayed progenitor (Badenes et al. 2009).

SNR 0509-67.5 already has a decisive result, where the entire central region is empty of any point source to V=26.9 ($M_V$=+8.4), so that all published single-degenerate models are eliminated, and only the double-degenerate model remains. While the SNR 0519-69.0 limits are not as tight (because the supernova is older and the remnant is in an area with much higher star density), nevertheless the result is important as it rules out all but two possible progenitor classes for this particular supernova, the supersoft sources and the double-degenerates.

\acknowledgments

The National Science Foundation supported this work through the REU Site in Physics \& Astronomy (NSF grant 10-04822) at Louisiana State University and through grant AST 11-09420.

{}

\begin{deluxetable}{lllll}
\tabletypesize{\scriptsize}
\tablewidth{0pc}
\tablecaption{Positions in SNR 0519-69.0}
\tablehead{\colhead{Position} & \colhead{RA (J2000)} & \colhead{Dec. (J2000)} & \colhead{Radius ($\arcsec$)} & \colhead{Confidence}}
\startdata
Geometric center in H$\alpha$	&	 05:19:34.80	&	 -69:02:06.46	&	0.7	&	1-$\sigma$	\\
Geometric center in X-ray	&	 05:19:34.87	&	 -69:02:07.38	&	0.7	&	1-$\sigma$	\\
Combined geometric center of SNR	&	 05:19:34.83	&	 -69:02:06.92	&	0.9	&	1-$\sigma$	\\
Site of explosion	&	 05:19:34.83	&	 -69:02:06.92	&	1.3	&	1-$\sigma$	\\
Red Giant ex-companion	&	 05:19:34.83	&	 -69:02:06.92	&	4.3	&	3-$\sigma$	\\
Subgiant ex-companion	&	 05:19:34.83	&	 -69:02:06.92	&	4.5	&	3-$\sigma$	\\
Main sequence ex-companion	&	 05:19:34.83	&	 -69:02:06.92	&	4.7	&	3-$\sigma$	\\
\enddata
\label{Table1}
\end{deluxetable}

\begin{deluxetable}{lllllll}
\tabletypesize{\scriptsize}
\tablewidth{0pc}
\tablecaption{Stars inside central error circle for SNR 0519-69.0}
\tablehead{\colhead{ID} & \colhead{RA (J2000)} & \colhead{Dec. (J2000)} & \colhead{$\Theta$ ($\arcsec$)} & \colhead{V (mag)} & \colhead{V-$H\alpha$ (mag)} & \colhead{Comments}}
\startdata
RG	 & 	 05:09:34.521	 & 	 -69:02:12.70	 & 	6	 & 	19.13	 $\pm$ 	0	 & 	0.47	 $\pm$ 	0	 & 	Nearest Red Giant	 \\
SG	 & 	 05:09:33.680	 & 	 -69:02:10.97	 & 	7.4	 & 	20.78	 $\pm$ 	0.01	 & 	0.35	 $\pm$ 	0.01	 & 	Nearest possible subgiant	 \\
1	 & 	 05:09:34.261	 & 	 -69:02:05.74	 & 	3.3	 & 	19.69	 $\pm$ 	0	 & 	-0.01	 $\pm$ 	0.01	 & 	Brightest star in circle	 \\
2	 & 	 05:09:34.850	 & 	 -69:02:06.83	 & 	0.1	 & 	23.37	 $\pm$ 	0.02	 & 	0.37	 $\pm$ 	0.04	 & 	\ldots	 \\
3	 & 	 05:09:34.787	 & 	 -69:02:06.69	 & 	0.4	 & 	25.67	 $\pm$ 	0.15	 & 	\ldots			 & 	\ldots	 \\
4	 & 	 05:09:34.801	 & 	 -69:02:07.27	 & 	0.4	 & 	25.22	 $\pm$ 	0.1	 & 	0.68	 $\pm$ 	0.16	 & 	\ldots	 \\
5	 & 	 05:09:34.884	 & 	 -69:02:07.57	 & 	0.7	 & 	23.97	 $\pm$ 	0.03	 & 	0.56	 $\pm$ 	0.06	 & 	\ldots	 \\
6	 & 	 05:09:34.967	 & 	 -69:02:06.82	 & 	0.7	 & 	23.47	 $\pm$ 	0.03	 & 	0.43	 $\pm$ 	0.05	 & 	\ldots	 \\
7	 & 	 05:09:34.954	 & 	 -69:02:07.32	 & 	0.8	 & 	22.09	 $\pm$ 	0.01	 & 	0.16	 $\pm$ 	0.02	 & 	V$<$22.7 MS star	 \\
8	 & 	 05:09:34.791	 & 	 -69:02:07.80	 & 	0.9	 & 	20.54	 $\pm$ 	0.01	 & 	0.04	 $\pm$ 	0.01	 & 	V$<$22.7 MS star	 \\
\enddata
\label{Table2}
\end{deluxetable}

\begin{figure}
\centering
\epsscale{.70}
\plotone{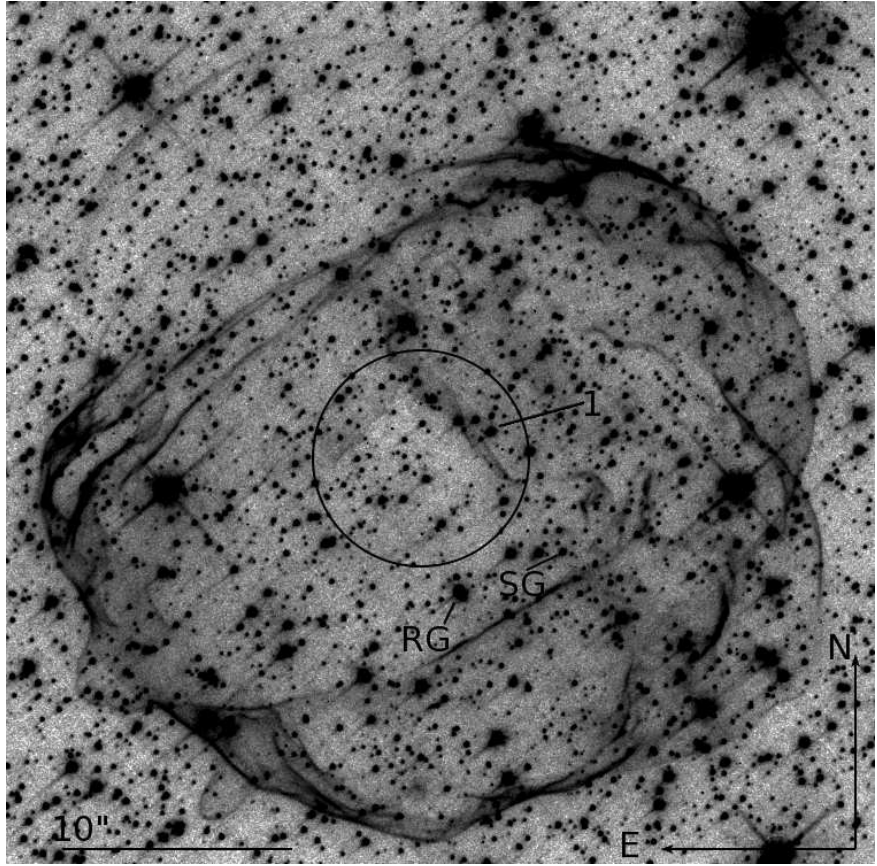}
\caption{The HST picture (a combination of the V-band and H$\alpha$ images) of SNR 0519-69.0 is shown here, with a circle marking the central region of the remnant. The center of this region is calculated from the entire edge of the shell as viewed in H$\alpha$ and X-ray light.  Note that the northeast quadrant of the shell has a faint circular arc outside the prominent arc cutting across the inside of the remnant.  Any ex-companion star can be offset from the center due to its original orbital velocity, a kick from the supernova, and ordinary measurement uncertainties in positioning the center, with a 4.7$\arcsec$ radius circle containing all possible ex-companions at the 99.73\% containment probability level.  This central region contains 27 main sequence stars brighter than $V=22.7$, and any one of these could be an ex-companion star from a supersoft source progenitor system.  The central region does not contain any red giants or subgiants, and this eliminates all the single-degenerate models (including the symbiotic, recurrent nova, helium donor star, and spin-up/spin-down models) that require a post-main sequence companion.  Only two published models remain, so the Type Ia supernova that created SNR 0519-69.0 must have come from either a double-degenerate system or a supersoft source. The star marked 1 is the brightest main sequence star in the central region, and the nearest red giant and subgiant (both outside the central region) are labeled RG and SG, respectively.}
\label{fig:zoom}
\end{figure}	

\begin{figure}
\centering
\epsscale{.70}
\plotone{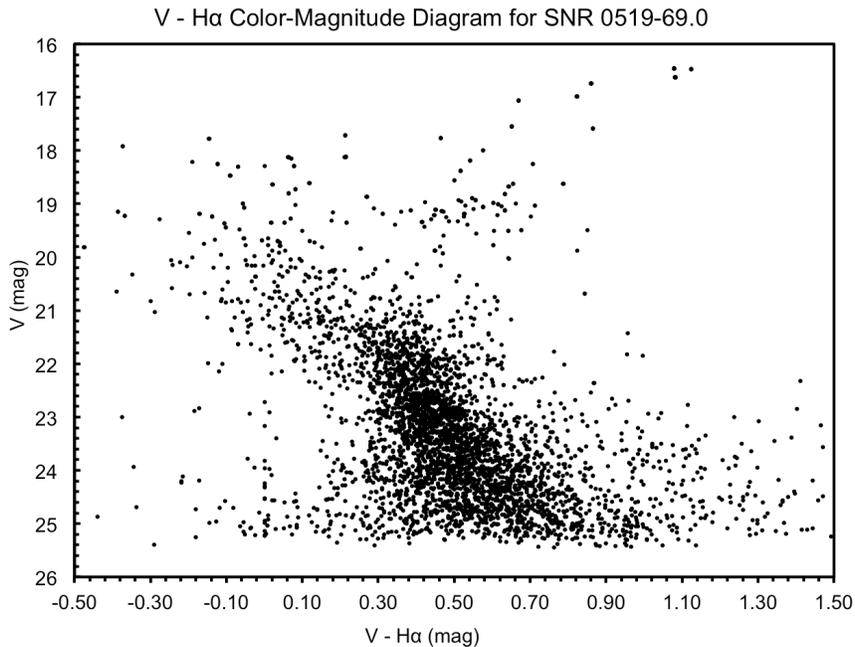}
\caption{This $V-H\alpha$ color-magnitude diagram can be used to identify post-main sequence stars in the field of SNR 0519-69.0. The $V-H\alpha$ color index, constructed with the currently-available HST images, is slightly unusual, but still shows a clear main sequence, red clump, and red giant branch.}
\label{fig:cmd}
\end{figure}

\end{document}